\documentclass[a4paper,11pt]{article}
\usepackage{pos}

\title{Far-from-equilibrium attractors in kinetic theory with two different relaxation times}
\ShortTitle{Far-from-equilibrium attractors in kinetic theory in 2RTA}

\author*[a]{Ferdinando Frasc\`{a}}
\author[a]{Andrea Beraudo}
\author{Michael Strickland}

\affiliation[a]{INFN, Sezione di Torino, via Pietro Giuria 1, I-10125 Torino}

\emailAdd{ferdinando.frasca@to.infn.it}
\emailAdd{beraudo@to.infn.it}
\emailAdd{michael.strickland@gmail.com}

\abstract{We solve a Boltzmann equation for massless quark and gluon fluids in a transversally homogeneous, longitudinally boost-invariant expansion. Quarks can be out of chemical equilibrium and the relaxation times of the two species are assumed to be connected by Casimir scaling. We numerically calculate moments of the distribution functions, identifying their early- and late-time attractors and reconstructing also the full distributions. These attractors appear when the system is still far from local thermalization, before hydrodynamics traditionally would be expected to apply. We also analyze the evolution of entropy production for different initial momentum anisotropies and quark abundances.}

\FullConference{10th International Conference on Quarks and Nuclear Physics (QNP2024)\\
8-12 July, 2024\\
Barcelona, Spain\\}

\begin{document}
\maketitle

\section{Introduction and general setup}

In heavy-ion collisions at the highest energies the formed matter is initially dominated by gluons, with quarks and antiquarks being pair produced afterwards. This results in a medium with nearly zero baryochemical potential, starting out of chemical equilibrium due to the excess of gluons and the lack of quarks/antiquarks~\cite{Biro:1993qt}. The deviation from chemical equilibrium is quantified by the quark fugacity $\gamma_q$ and quarks are assumed to be massless, for the sake of simplicity.

We investigate the existence of dynamical attractors in kinetic theory when quarks and gluons have different time-dependent relaxation times, related via Casimir scaling. This leads to the following Landau matching condition~\cite{Florkowski:2012as}:
\begin{equation}
    \label{eq:LandauMC}
    \varepsilon_q + \frac{\varepsilon_g}{C_R} = \varepsilon_{q, {\rm eq}} + \frac{\varepsilon_{g, {\rm eq}}}{C_R}\,,\,\,\, {\rm where} \,\,\, \varepsilon_{g, {\rm eq}} = \frac{3 \hspace{0.07cm} g_g}{\pi^2} \hspace{0.07cm} T^4 \hspace{0.3cm} {\rm and} \hspace{0.3cm}  \varepsilon_{q, {\rm eq}} = \frac{6 \hspace{0.07cm} g_q}{\pi^2} \hspace{0.07cm} T^4\,,
\end{equation}
allowing one to define an effective temperature for our out-of-equilibrium medium. In Eq.~(\ref{eq:LandauMC}) $C_R \equiv \tau_{\rm eq, g}/\tau_{\rm eq, q} = 4/9$ and $g_g = 16$, $g_q = 18$ count the internal degrees of freedom of the two species.

In the case of Bjorken flow ($v_z = z/t$), the Boltzmann equation (BE) in relaxation time approximation (RTA) for the on-shell one-particle distribution reads
\begin{equation}
    \partial_{\tau} \hspace{0.05cm} f_a = -\frac{f_a-f_{{\rm eq}, a}}{\tau_{{\rm eq}, a}} \,, \,\,\, \text{ \small{with}} \hspace{0.3cm} a = q, g\,\,\, {\rm and} \,\, \tau \equiv \sqrt{t^2 - z^2}\,.
\end{equation}
The above equation, similarly to the case of a single relaxation time, can be solved exactly:
\begin{equation}
    \label{BF12}
    f_a (\tau; w, p_T) = D_a (\tau, \tau_0) \hspace{0.07cm} f_{0, a} (w, p_T) + \int_{\tau_0}^{\tau} \frac{d \tau'}{\tau_{{\rm eq}, a}(\tau')} \hspace{0.07cm} D_a (\tau, \tau') \hspace{0.07cm} f_{{\rm eq}, a} (\tau'; w, p_T)\,,
\end{equation}
with $w \equiv t \hspace{0.07cm} p_z - z \hspace{0.07cm} E$ being an invariant quantity  under longitudinal (along the $z$-axis) boosts. In the above equation the damping function $D_a(\tau_2,\tau_1) \equiv \exp \left[ - \int_{\tau_1}^{\tau_2}\frac{d \tau'}{\tau_{{\rm eq}, a}(\tau')} \right]$ was introduced, which quantifies the fraction of particles which have not undergone any interaction during the time interval from $\tau_1$ to $\tau_2$~\cite{Broniowski:2008qk}. We choose the Strickland-Romatschke form to initialize the one-particle distributions~\cite{Alqahtani:2017mhy}:
\begin{equation}
    \label{rta3}
    f_{0,a}(x;p) \equiv G_{0, a} \hspace{0.07cm} \exp \left[ - {\sqrt{p_T^2 + (1 + \xi_0) \hspace{0.07cm} p_z^2}}/{\Lambda_0}  \right]\,,
\end{equation}
where $G_{0, g} = g_g$ and $G_{0, q} = 2 \hspace{0.07cm} g_q \hspace{0.07cm} \gamma_{q, 0}$ (the factor 2 accounts for the particle-antiparticle degeneracy). Furthermore $\Lambda_0$ represents the transverse-momentum scale of the particles, which reduces to the temperature $T_0$ if the system is in equilibrium, i.e. $\gamma_q = 1$ (chemical) and $\xi_0 = 0$ (kinetic).

In the following we display some numerical results obtained within the above setup. For a more comprehensive discussion we refer the reader to Ref.~\cite{Frasca:2024ege}.

\section{Evolution of the single-particle distributions and quark abundance}

We consider the fluid's local rest frame, where the distribution function becomes isotropic when equilibrium is reached. Exploiting longitudinal boost invariance we focus on the fireball center, setting $z = 0$. We then show the evolution of $f_a$ with the scaled time $\overline{\omega} \equiv \tau/\tau_{\rm eq}$, where $\tau_{\rm eq}$ is the gluon relaxation time. To initialize the system we set $\tau_0 = 0.15$ fm/c, $T_0 = 600$ MeV and $\eta/s = 0.2$.

\begin{figure}[!hbt]
    \centering
    \includegraphics[width=0.332\textwidth] {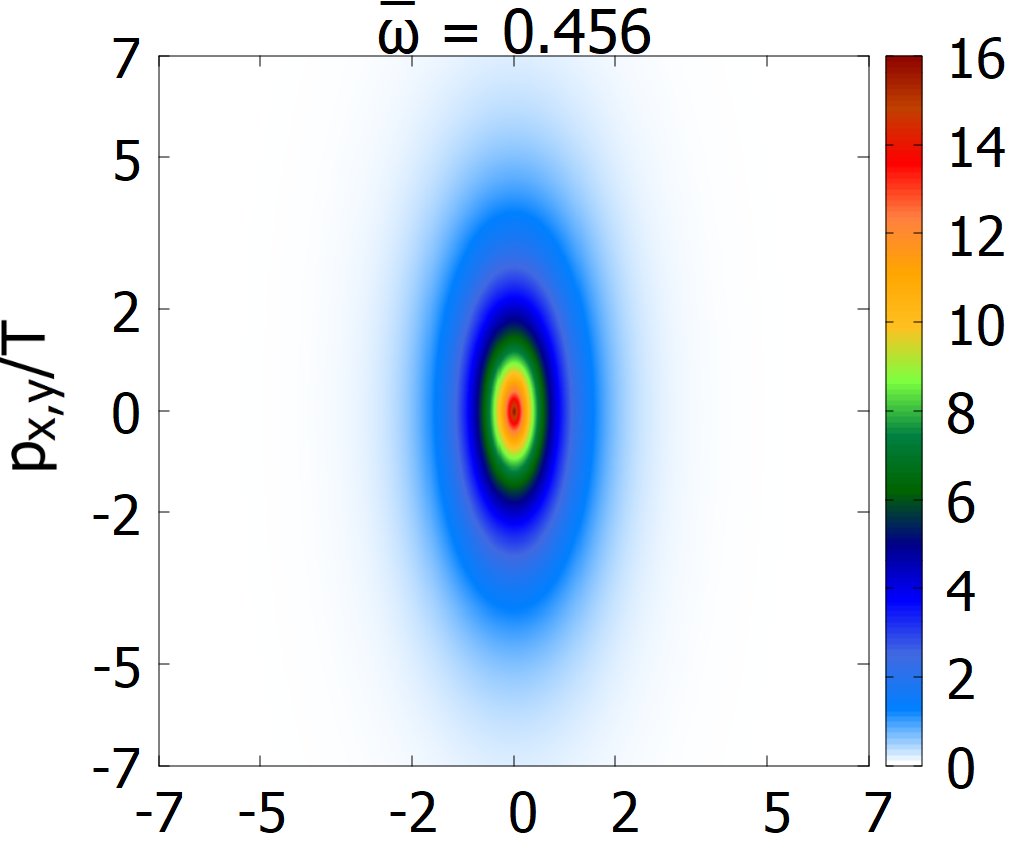} 
    \includegraphics[width=0.305\textwidth] {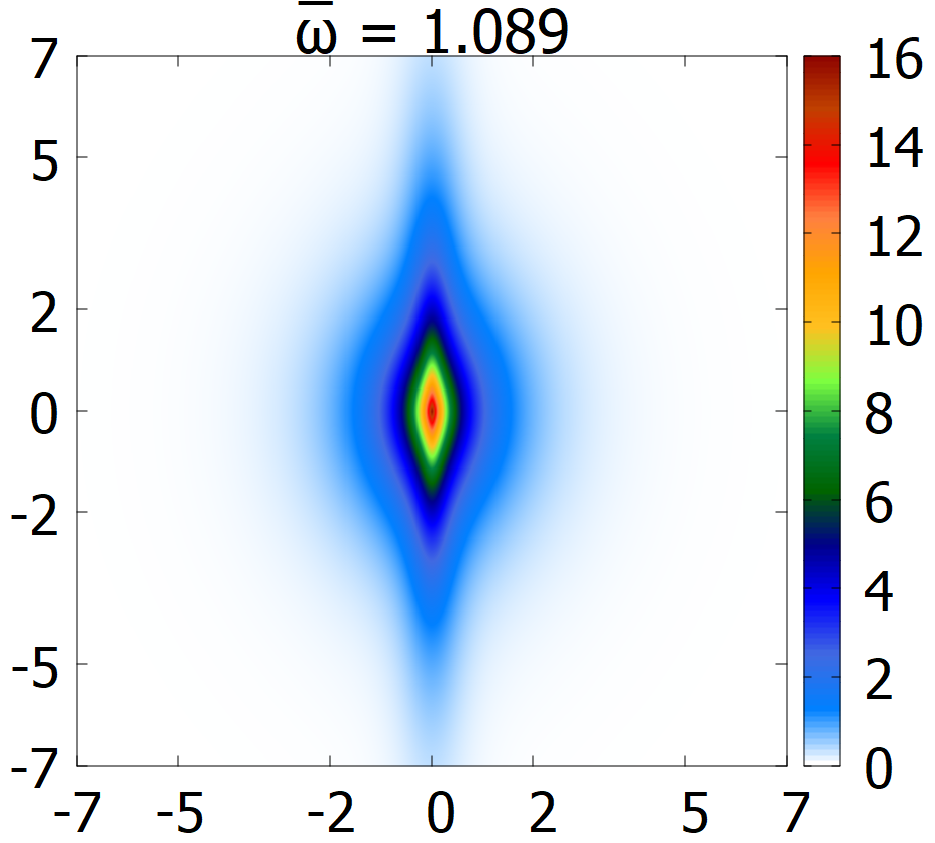}
    \includegraphics[width=0.305\textwidth] {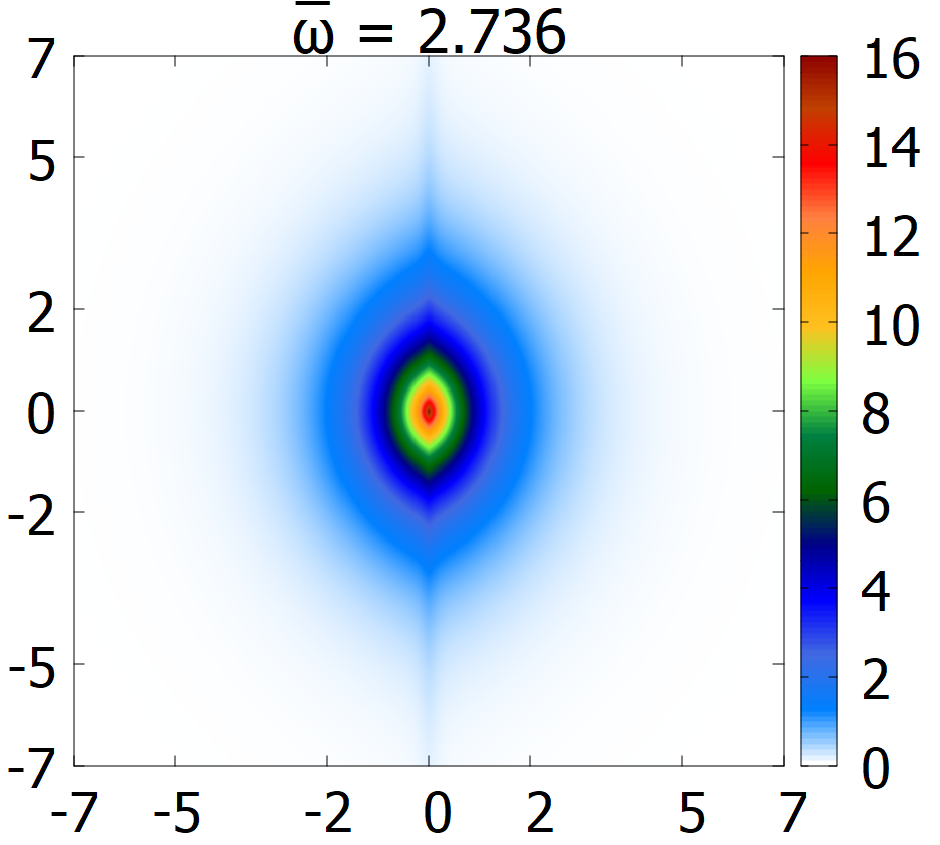} \\
    \vspace{0.1cm}
    \includegraphics[width=0.335\textwidth] {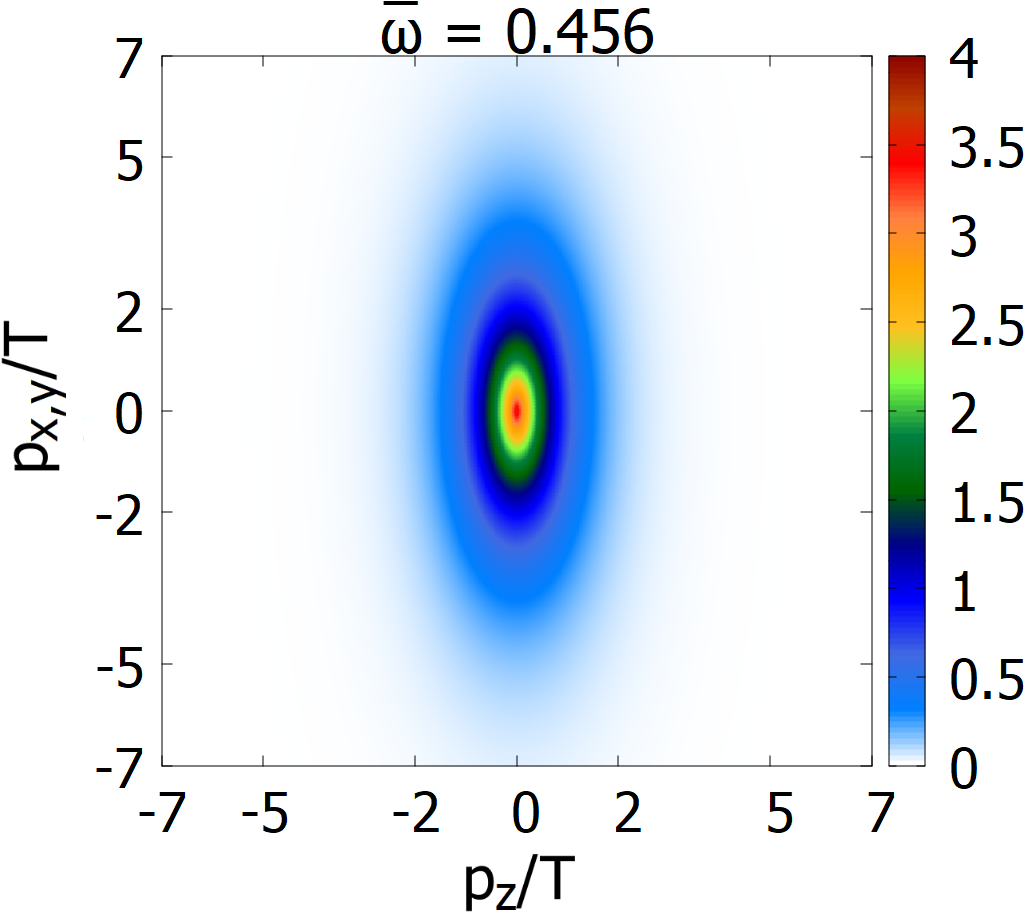} 
    \includegraphics[width=0.305\textwidth] {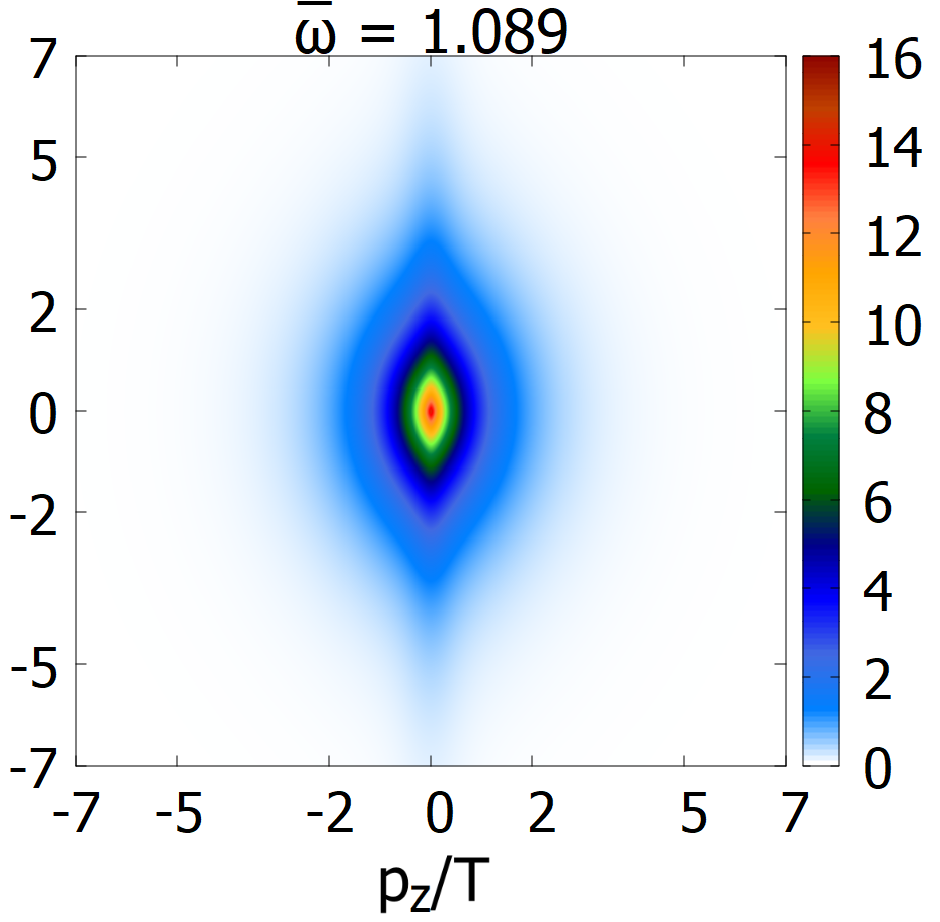}
    \includegraphics[width=0.305\textwidth] {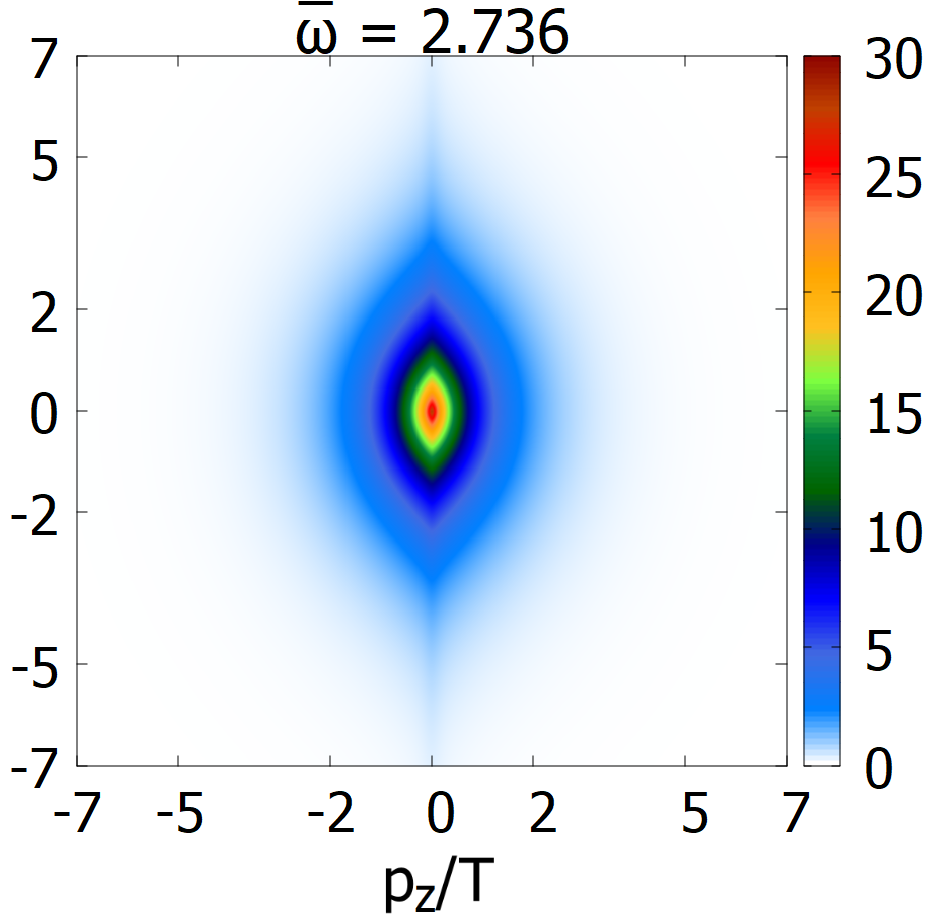}  
    \caption{Snapshots of the gluon (upper panels) and quark (lower panels) momentum distribution (2RTA-BE) for different
scaled times $\Bar{\omega}$. Here $\alpha_0 \equiv (1 + \xi_0)^{-1/2} = 0.4$ and $\gamma_{q,0}=0.1$ (gluon-dominated plasma).}
    \label{f_gq}
\end{figure}

In Fig.~\ref{f_gq}, two distinct distribution components are evident. Non-interacting particles create an anisotropic distribution along $p_z \sim 0$ (non-hydrodynamic modes). This free-streaming contribution, described by the first term in Eq.~(\ref{BF12}), would increase anisotropy, but its amplitude decays exponentially due to the damping function $D_a$, leading to hydrodynamization. The second component, the ``evolution term'' in Eq.~(\ref{BF12}), causes isotropization and dominates when $\overline{\omega} > 1$, as particle interactions overwhelm plasma expansion. Isotropization is slower for quarks than for gluons due to their longer relaxation time $\tau_{\rm eq, q} = \tau_{\rm eq}/C_R$.

\section{Results for the moments of the distribution functions}

\begin{figure}[!hbt]
    \centering
    \includegraphics[width=0.325\textwidth] {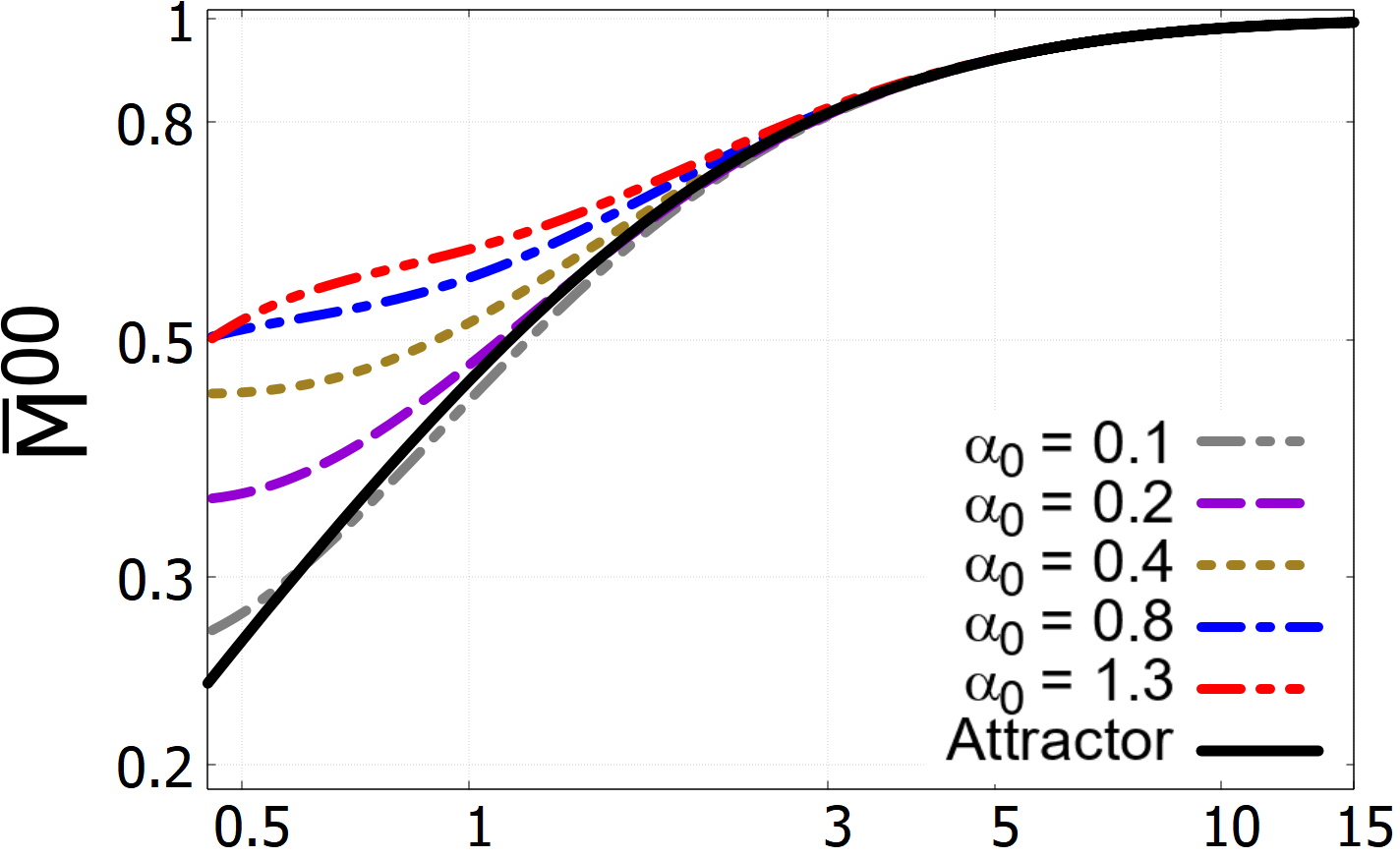} 
    \includegraphics[width=0.325\textwidth] {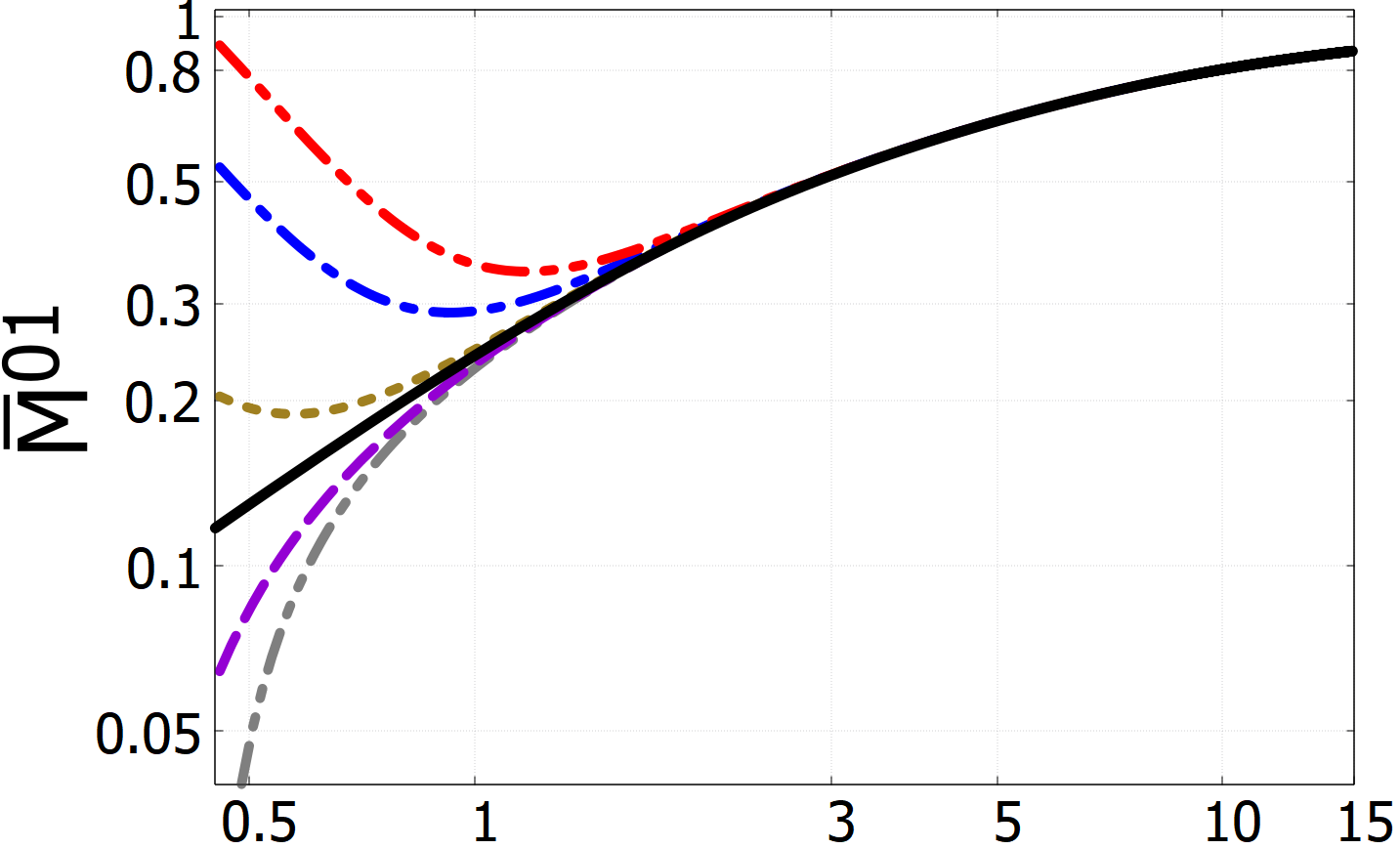}
    \includegraphics[width=0.325\textwidth] {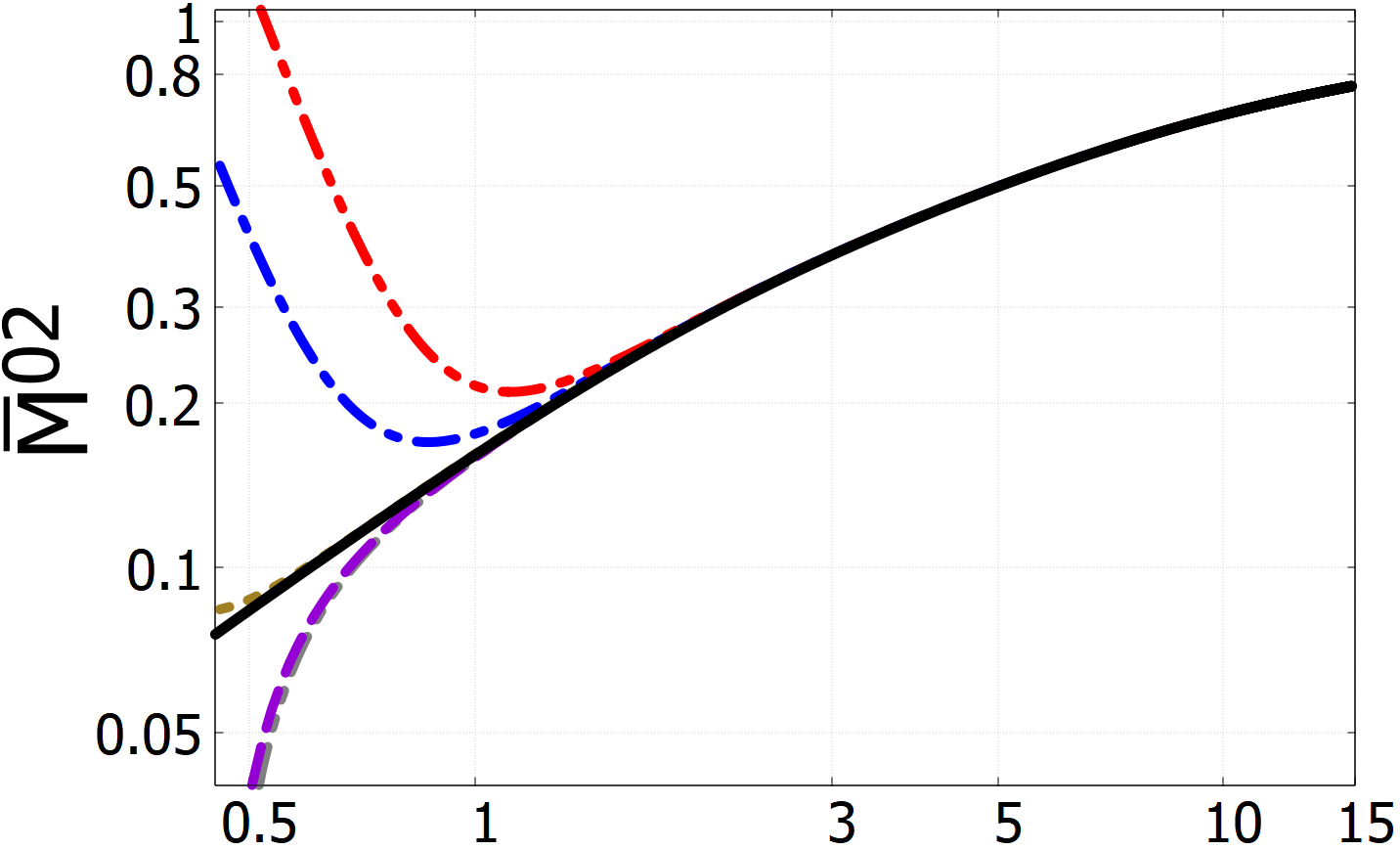} \\
    \includegraphics[width=0.325\textwidth] {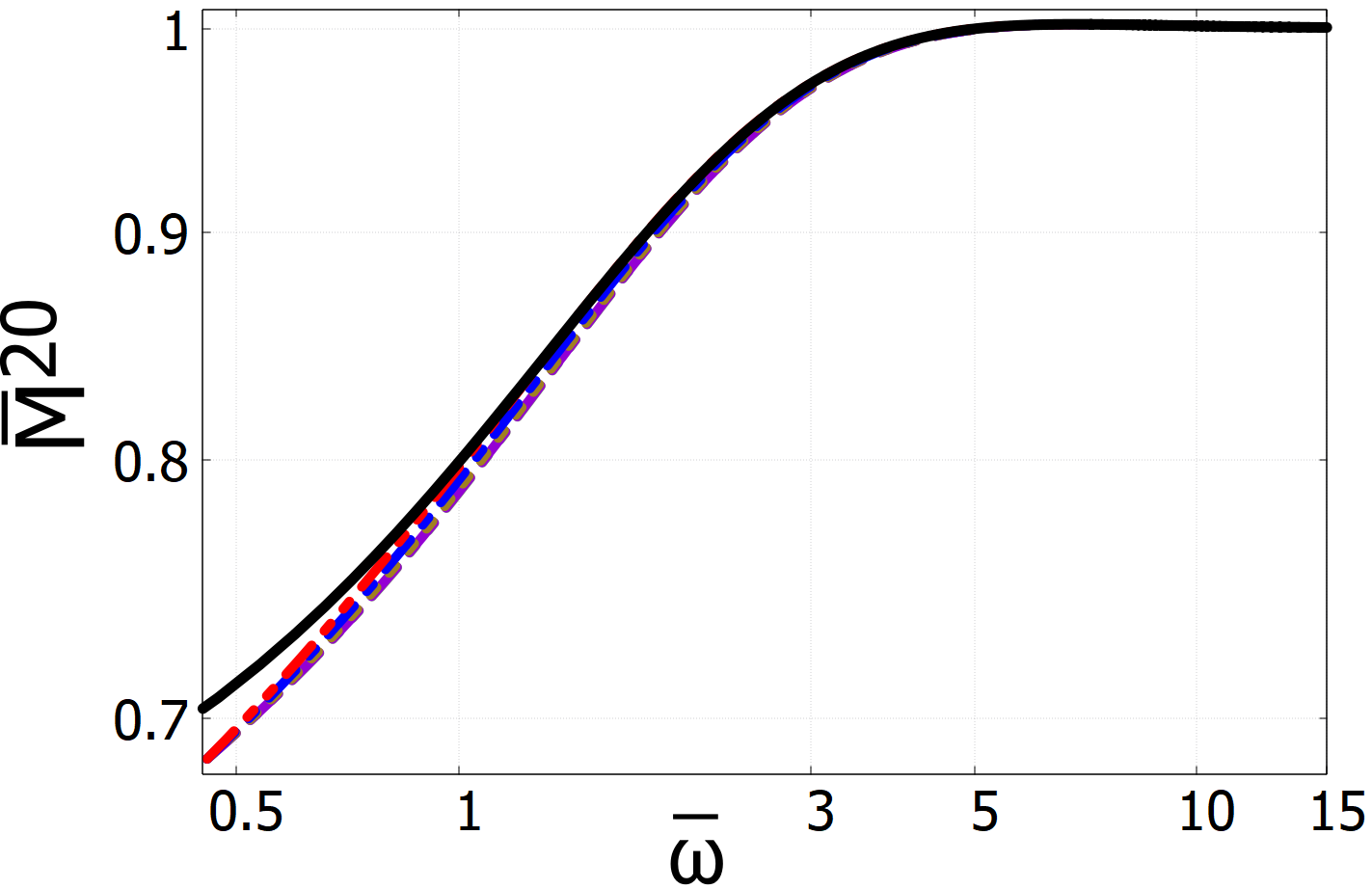} 
    \includegraphics[width=0.325\textwidth] {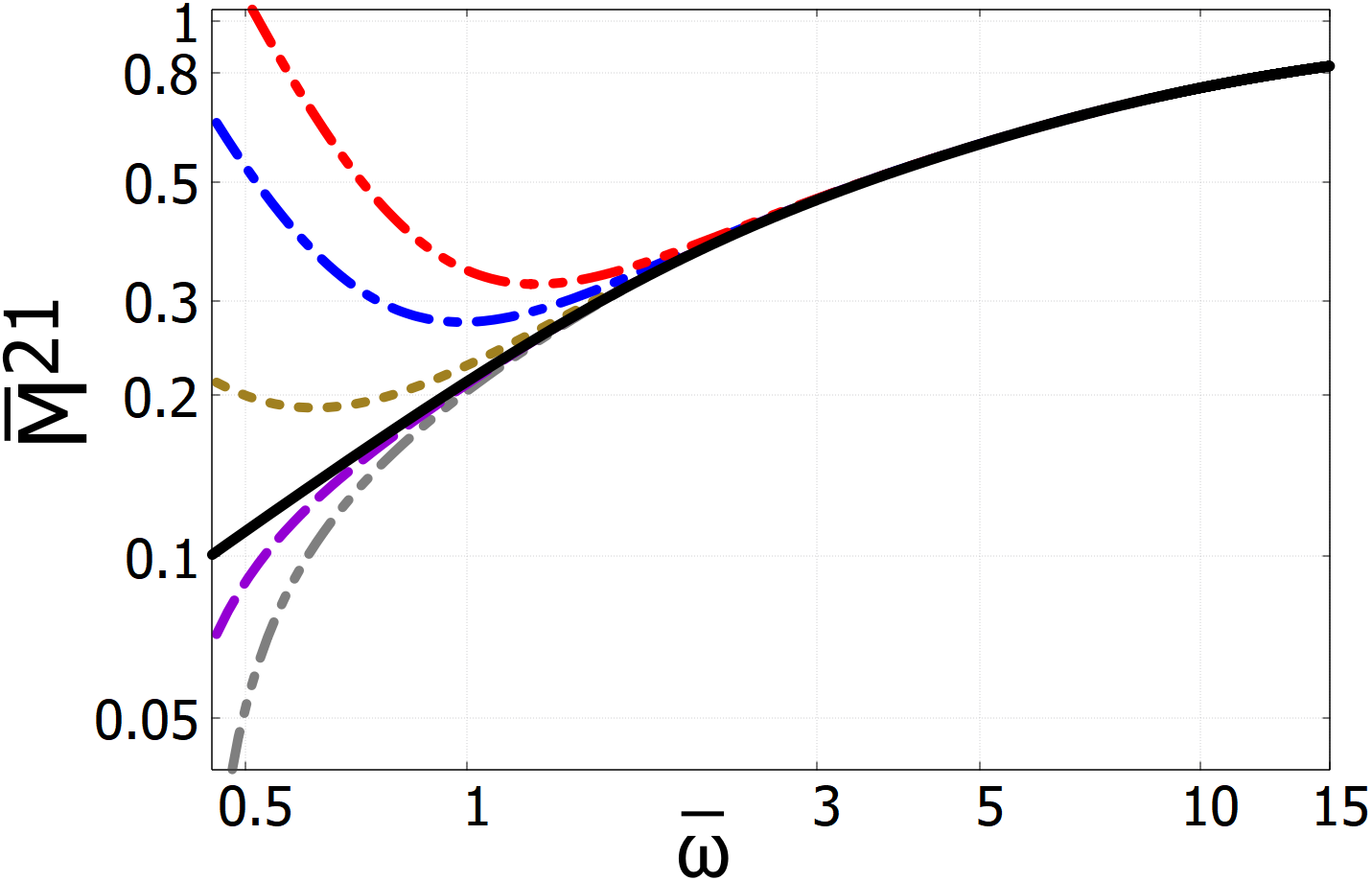}
    \includegraphics[width=0.325\textwidth] {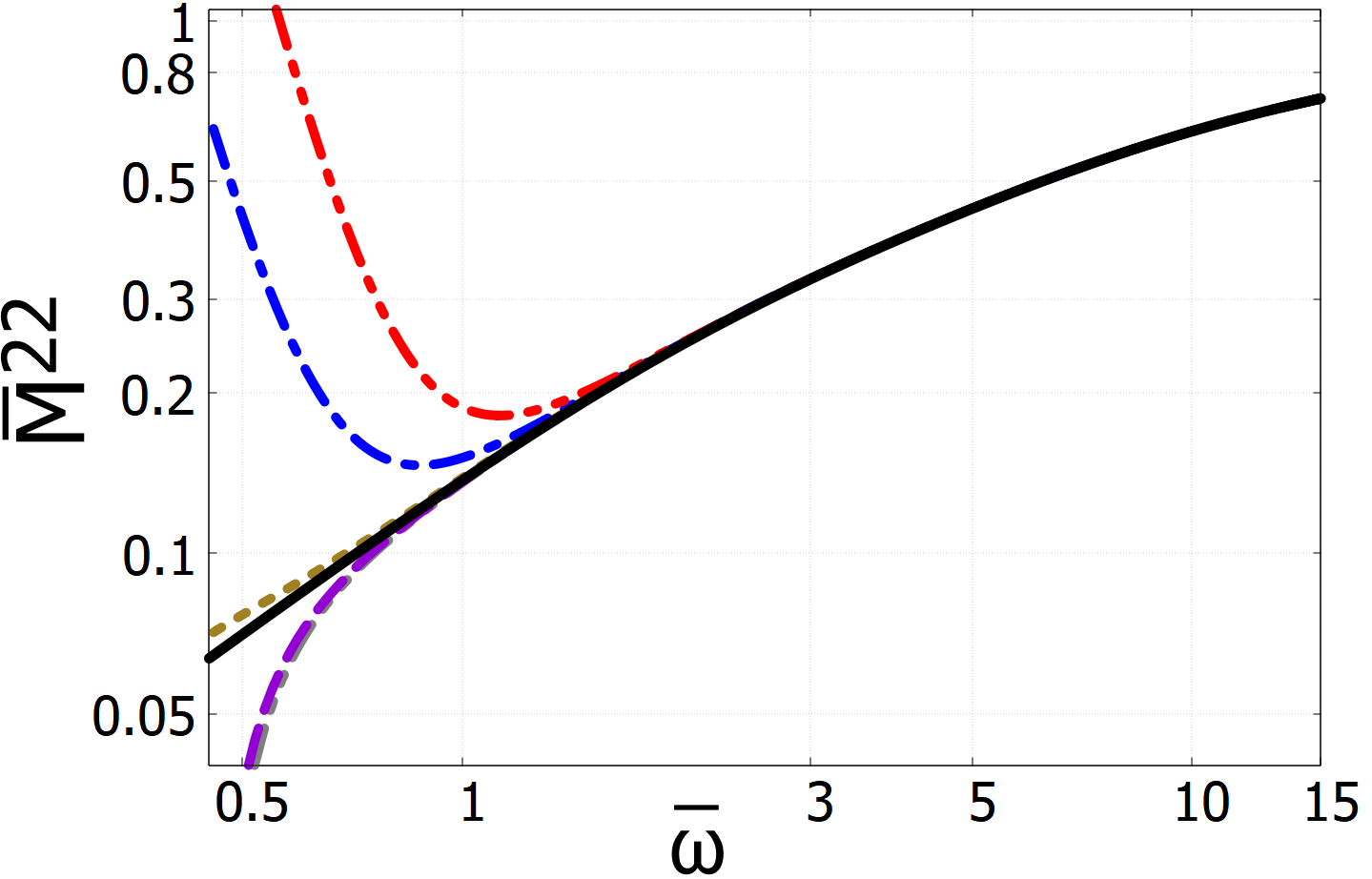}  
    \caption{Scaled moments $\overline{M}^{\hspace{0.05cm} nm}$ obtained from the late-time attractor solution (solid black line) compared to a set of exact solutions {of the 2RTA-BE} (various dashed colored lines).}
    \label{fig:mom-late-under}
\end{figure}

\begin{figure}[!hbt]
    \centering
    \includegraphics[width=0.325\textwidth] {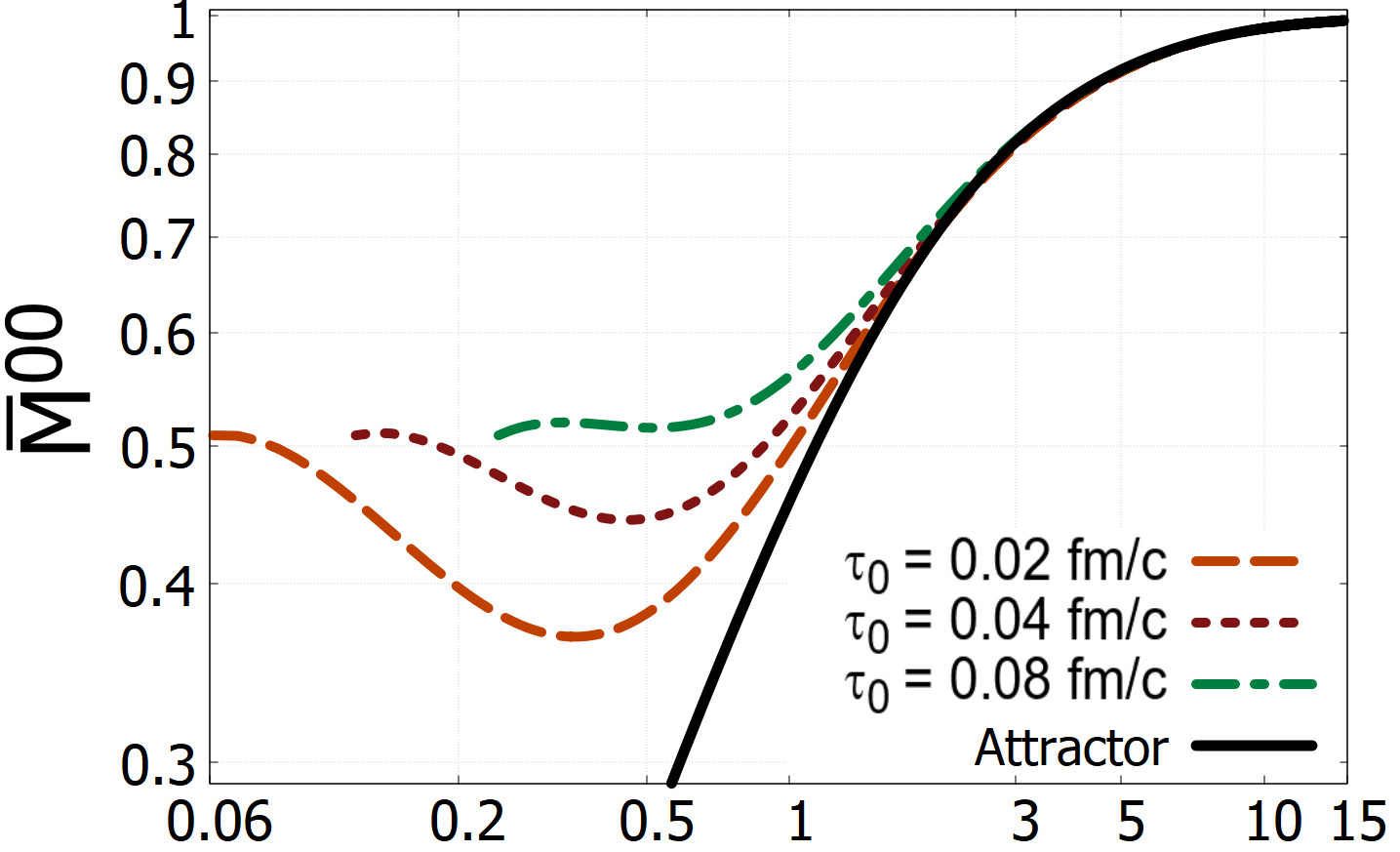} 
    \includegraphics[width=0.325\textwidth] {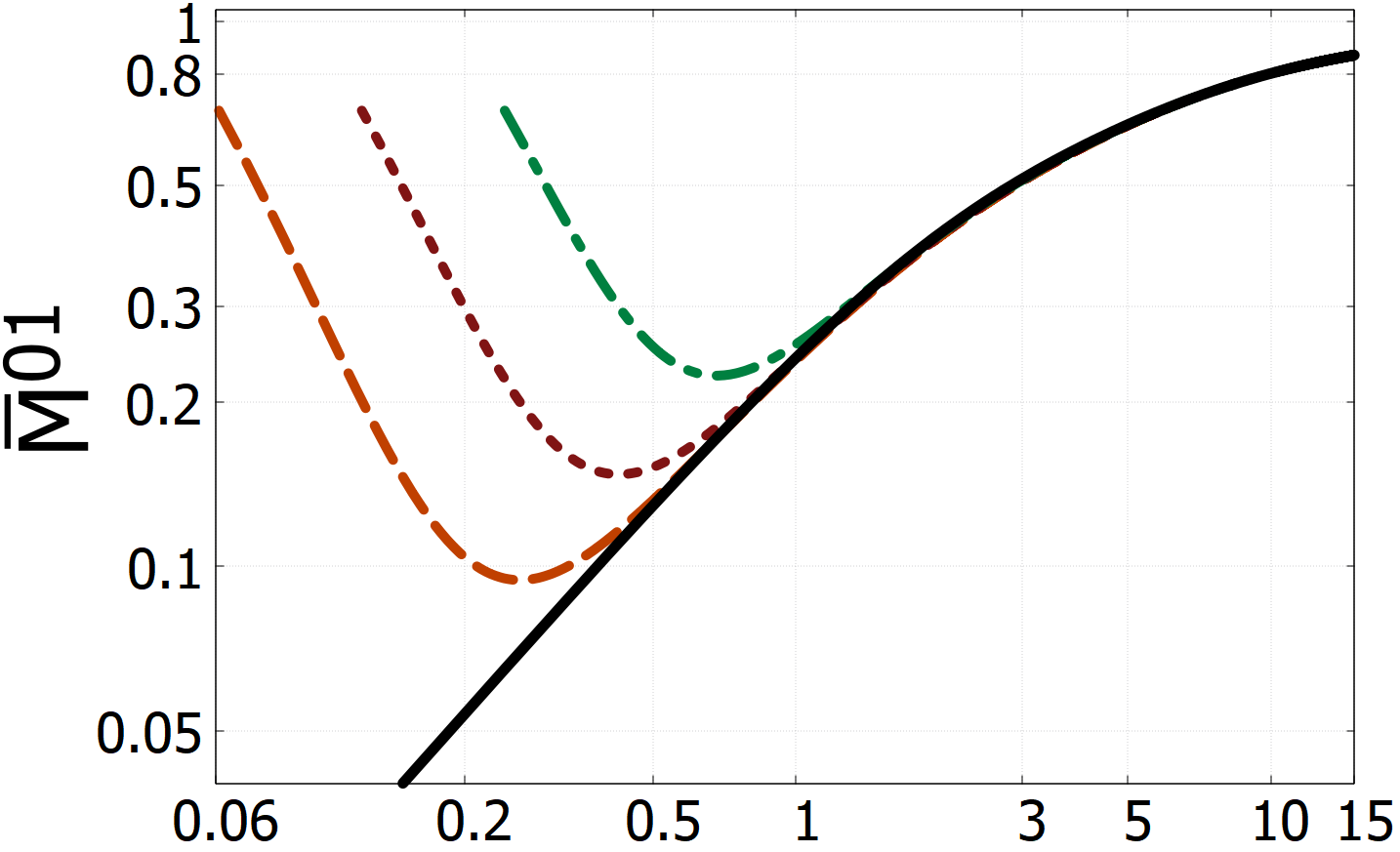}
    \includegraphics[width=0.325\textwidth] {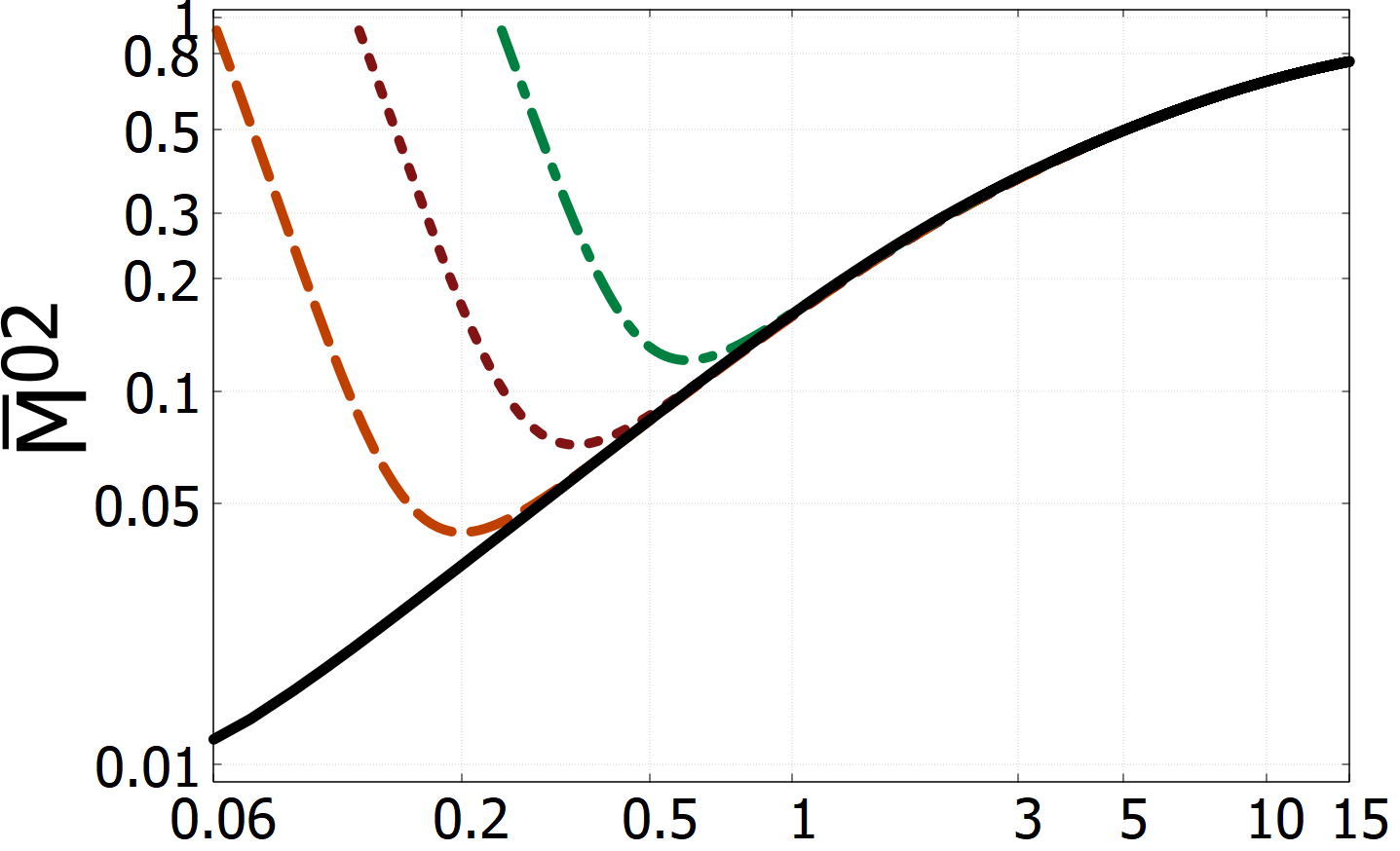} \\
    \includegraphics[width=0.325\textwidth] {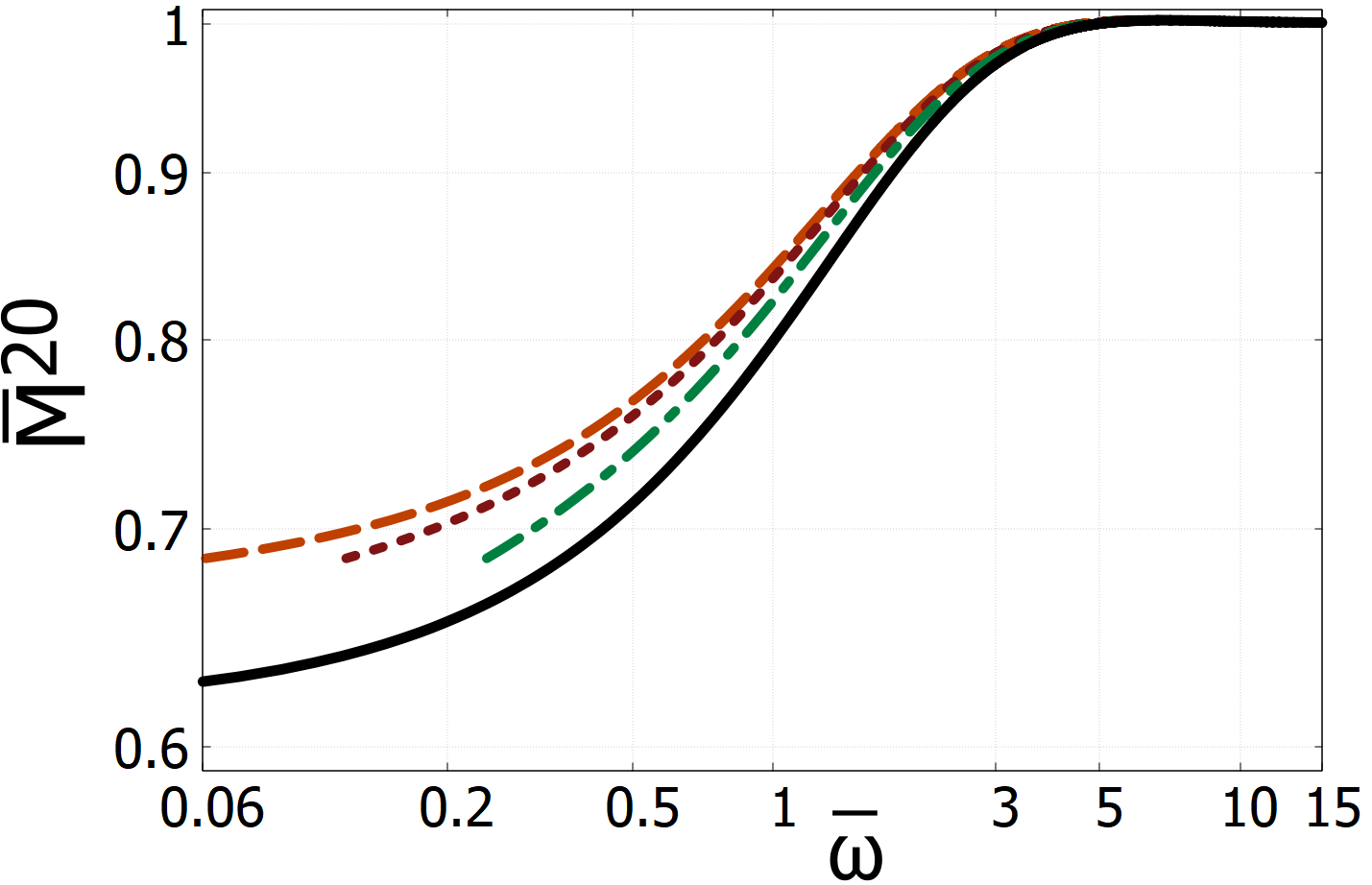} 
    \includegraphics[width=0.325\textwidth] {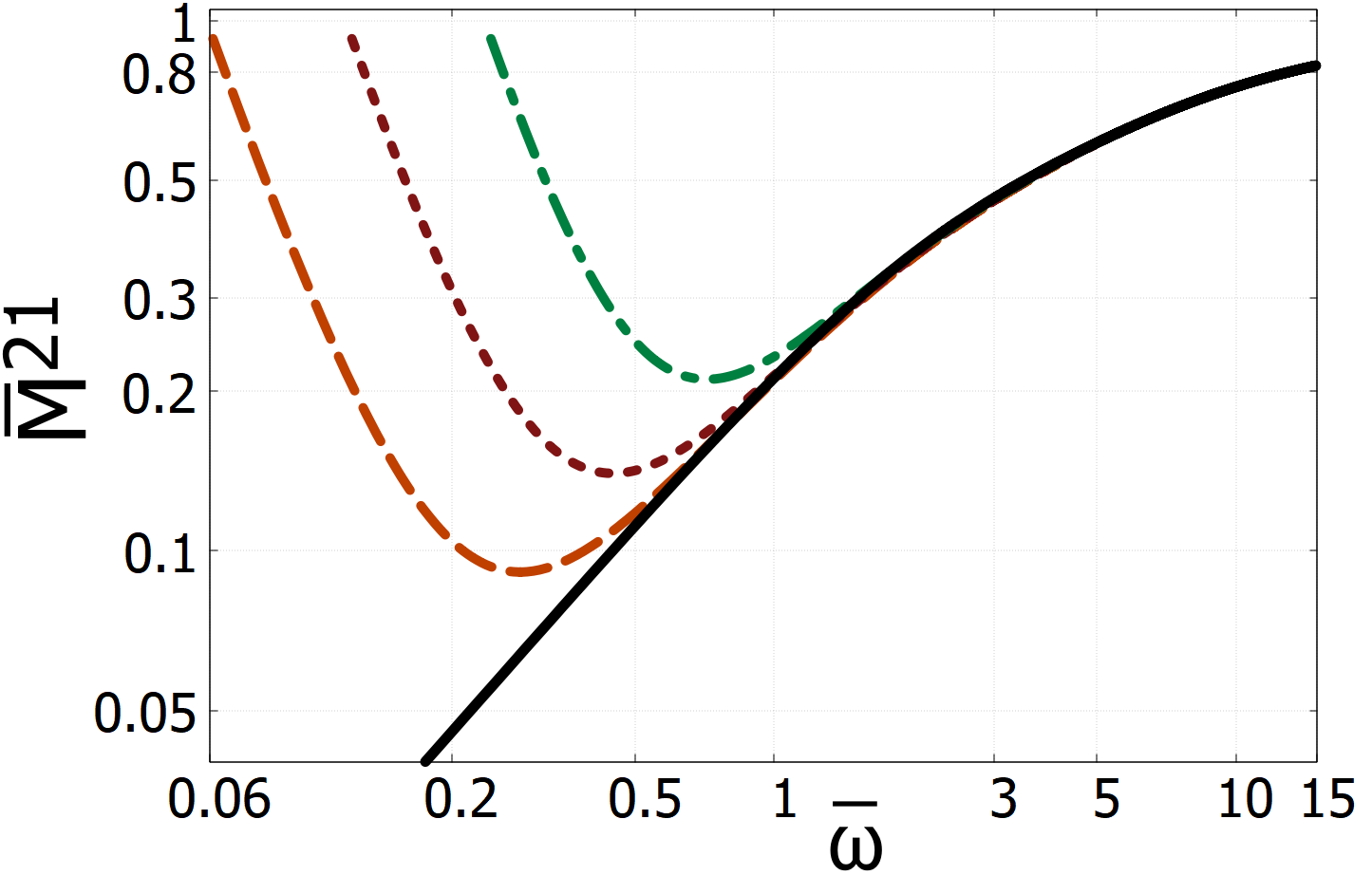}
    \includegraphics[width=0.325\textwidth] {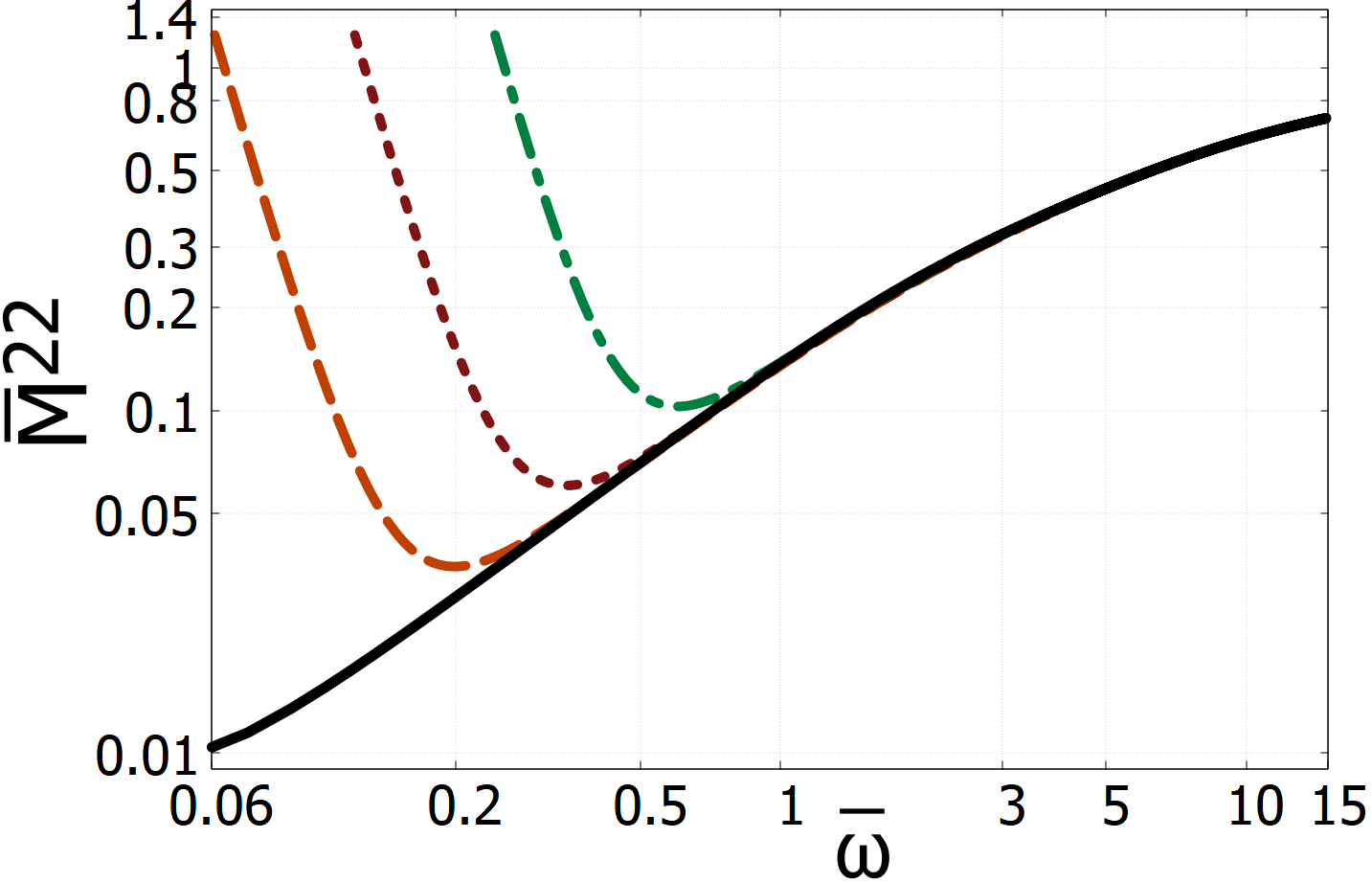}  
    \caption{Study of the convergence to the early-time attractor (solid black line) of the scaled moments $\overline{M}^{\hspace{0.05cm} nm}$ (various dashed colored lines) obtained from the exact solution of the 2RTA-BE.}
    \label{earl1}
\end{figure}

We define the general moments associated with the distribution of the species $a$ in the case of a purely longitudinal expansion with $u^\mu\equiv(t/\tau,0,0,z/\tau)$ and $z^\mu\equiv(z/\tau,0,0,t/\tau)$ as~\cite{Strickland:2018ayk}:
\begin{equation}
\label{robo}
    M_a^{nm} \equiv \int d \chi \hspace{0.07cm} ( p \cdot u)^n \hspace{0.07cm} ( p \cdot z)^{2m} \hspace{0.07cm} f_a( \tau; w, p_T)\,, 
\end{equation}
where $f_a$ corresponds to the exact solution of the 2RTA-BE in Eq.~(\ref{BF12}). In the above the invariant integration measure for a system of on-shell massless particles can be rewritten as follows
\begin{equation*}
 d \chi \equiv  \frac{d^2 p_T\, dw}{(2\pi)^3 \,v}\,,\quad{\rm with}\quad  v = \sqrt{w^2 + p^2_T \hspace{0.07cm} \tau^2}\,.
\end{equation*}
The integral equation, describing the evolution of the total moment $M^{nm} \equiv \sum_a M^{nm}_a$, can be more conveniently expressed as the sum of two contibutions:
\begin{equation}
    \label{eq:mom-split}
        M^{nm}(\tau) = M^{nm}_{0}(\tau) + M^{nm}_{\rm coll}(\tau)\,.
\end{equation}
The first term comes from the free-streaming contribution in Eq.~(\ref{BF12}) and reads~\cite{Frasca:2024ege}
\begin{multline}\label{eq:split-fs}    
    M^{nm}_{0}(\tau)\! \equiv\! A \left\{D(\tau, \tau_0) \Bigl[ r \!+\! 2  \,\gamma_{q, 0} \, \bigl( D(\tau,\tau_0)\bigr)^{C_R - 1} \Bigr] \,
    \biggl( \frac{2 \hspace{0.07cm} \left(2 + \Bar{r}\right)}{2 \,\gamma_{q, 0} + \Bar{r}} \biggr)^\frac{n + 2m + 2}{4} T_{0}^{n +2m +2} \,\frac{H^{nm}\left( \alpha_0 \frac{\tau_0}{\tau} \right)}{\Bigl[H(\alpha_0) \Bigr]^{\frac{n +2m +2}{4}}} \right\} \,,
\end{multline}    
while the second one contains the contribution of collisions between quarks and gluons:
\begin{align}
\begin{split}\label{eq:split-coll}
    M^{nm}_{\rm coll}(\tau) \!\equiv\! A \!\hspace{0.07cm} \Biggl\{C_R \hspace{0.07cm} \int_{\tau_0}^{\tau} \frac{d \tau'}{\tau_{\rm eq}(\tau')} \hspace{0.07cm} D(\tau, \tau') \hspace{0.07cm} \biggl[ \Bar{r} + 2 \hspace{0.07cm} \bigl( D(\tau,\tau')\bigr)^{C_R - 1} \biggr] \hspace{0.07cm}T^{\hspace{0.07cm} n +2m +2}(\tau') \hspace{0.07cm} H^{nm} \left( \frac{\tau'}{\tau} \right) \Biggr\}\,.
\end{split}
\end{align}
In the above we introduced two new constants. The first one is given by
\begin{equation}
\label{nonna}
    \Bar{r} \equiv \frac{r}{C_R} \,,\,\,\, \text{ \small{with}} \quad r \equiv \frac{g_g}{g_q} = \frac{8}{9}\,,
\end{equation}
while the second overall constant reads $A \equiv g_q \hspace{0.07cm} \frac{\Gamma(n + 2m +2)}{(2\pi)^2}$. Note that in Eq.~(\ref{eq:split-coll}) hypergeometric special functions appear:
\begin{equation}
    H^{nm}(y) = \frac{2}{2m + 1} \hspace{0.07cm} y^{2m + 1} \hspace{0.07cm} _2F_1 \left(m + \frac{1}{2}, \frac{1 - n}{2}, m + \frac{3}{2}; 1 - y^2 \right)\,.
\end{equation}
Eq.~(\ref{eq:mom-split}) can be numericaly solved only after having determined the evolution of the temperature~\cite{Frasca:2024ege}
\begin{align}
    \begin{split}
    \label{T_exff}
        T^4(\tau) &=  D(\tau, \tau_0) \hspace{0.07cm} \biggl[ \Bar{r} + 2 \hspace{0.07cm} \gamma_{q, 0} \hspace{0.07cm} \bigl( D(\tau,\tau_0)\bigr)^{C_R - 1} \biggr] \hspace{0.07cm} \bigl( 2 \hspace{0.07cm} \gamma_{q,0} + \Bar{r} \bigr)^{-1} \hspace{0.07cm} T_0^4 \hspace{0.07cm} \frac{H \left( \alpha_0 \hspace{0.07cm} \frac{\tau_0}{\tau} \right)}{H(\alpha_0)} \hspace{0.1cm}+ \\
        &+ \frac{C_R}{2 + \Bar{r}} \hspace{0.07cm} \int_{\tau_0}^{\tau} \frac{d \tau'}{2 \hspace{0.07cm} \tau_{\rm eq}(\tau')} \hspace{0.07cm} D(\tau, \tau') \hspace{0.07cm} \biggl[ \frac{\Bar{r}}{C_R} + 2 \hspace{0.07cm} \bigl( D(\tau,\tau')\bigr)^{C_R - 1} \biggr] \hspace{0.07cm} T^4(\tau') \hspace{0.07cm} H \left( \frac{\tau'}{\tau} \right) \,,
    \end{split}
\end{align}
which is a consequence of the Landau matching condition in Eq.~(\ref{eq:LandauMC}). It is convenient to introduce the so-called scaled moments $\overline{M}^{\hspace{0.05cm} nm}(\tau) \equiv \frac{M^{nm}(\tau)}{M^{nm}_{\rm eq}(\tau)}$, where
\begin{equation}
    \label{BF18}
    M_{\rm eq}^{nm}(\tau) = g_q \hspace{0.07cm} \frac{\Gamma(n + 2m +2)}{(2 \pi)^2} \hspace{0.07cm} \frac{2 \hspace{0.07cm} \bigl( 2 + r \bigr)}{2 \hspace{0.07cm} m + 1} \hspace{0.07cm} T^{n + 2m + 2}(\tau)\,.
\end{equation}
We { have} examined the late-time attractor { behavior} (Fig.~\ref{fig:mom-late-under}) by varying the initial anisotropy { coefficient} $\xi_0$, and the early-time attractors (Fig.~\ref{earl1}) { by changing} $\tau_0$ while keeping the initial conditions (the same as in Fig.~\ref{f_gq}) constant~\cite{Strickland:2017kux}.
Our numerical results reveal that each moment exhibits a late-time attractor developing before the system is close to local thermalization, except the scaled energy density $ \varepsilon/\varepsilon_{\rm eq} = \overline{M}^{\hspace{0.05cm} 20}$, where the concept of off-equilibrium attractor is less significant. However, only moments with $m > 0$ show an early-time attractor (convergence for $\overline{\omega} \lesssim 1$). In these cases, the dependence on $p_z$ -- see Eq.~(\ref{robo}) -- indicates that universality is achieved due to the system rapid longitudinal expansion in the early stages.

\section{Entropy density and entropy production in a Bjorken-expanding mixture}

Once the exact distribution functions for partons (here treated as classical particles) are known, one can compute the entropy density of the system via the following integral in momentum space~\cite{Florkowski:2013lya}:
\begin{equation}
    \label{entropy}
    s(\tau) = - \sum_{a = g, q} G_a(\tau) \hspace{0.07cm} \int d\chi \hspace{0.07cm} (p \cdot u) \hspace{0.07cm} \widehat{f}_a(\tau; w , p_T) \hspace{0.07cm} \biggl\{ \ln \Bigl[ \widehat{f}_a(\tau; w , p_T) \Bigr] - 1 \biggr\}\,.
\end{equation}
In Eq.~(\ref{entropy}), $\widehat{f}_a$ represents the functional part of the exact distribution function in Eq.~(\ref{BF12}), {the one carrying the dependence on the particle momentum}, without the number of active degrees of freedom, which are factorized out of the integral. Numerical simulations show a rapid entropy production, particularly in the pre-hydrodynamic phase. In fact, most entropy is generated before hydrodynamization, expected to occur when $\overline{\omega} \gtrsim 5$, where individual runs collapse onto the late-time attractor. Entropy production is measured using the $\frac{s\!\cdot\!\tau}{s_0\!\cdot\!\tau_0}$ ratio, in which the entropy density of our dissipative system is compared to the one of an ideal fluid, for which $s_{\rm id}(\tau) = s_0 \hspace{0.07cm} \frac{\tau_0}{\tau}$.

\begin{figure}[!hbt]
    \centering 
    \includegraphics[width=0.48\textwidth] {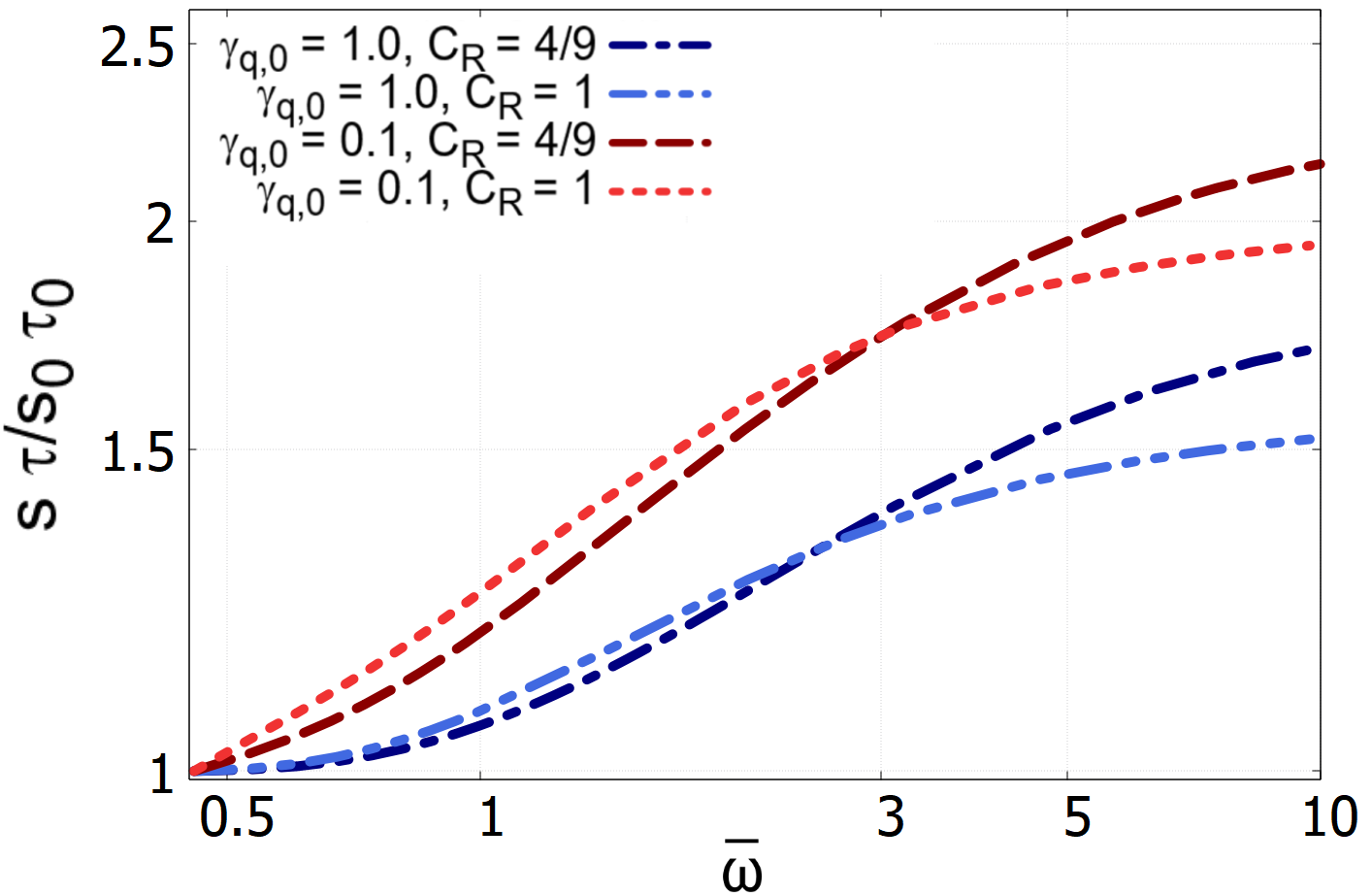}    
    \caption{{Dependence of the entropy production on the initial quark abundance ($\gamma_{q,0}=0.1$ vs $\gamma_{q,0}=1$) and on the different relaxation time of quarks and gluons ($C_R=4/9$ vs $C_R=1$)}. Here $\alpha_0 \equiv (1 + \xi_0)^{-1/2} = 0.8$.}
    \label{comp}
\end{figure}

From Fig.~\ref{comp}, we observe that having two relaxation times results in a modest increase in entropy production during the Bjorken expansion (+15\% by the end of the simulation). However, there is a much stronger dependence on the initial quark abundance, with a final 40\% increase in entropy.

\acknowledgments
F.F. and A.B. acknowledge financial support by MUR within the Prin$\_$2022sm5yas project.


\begin{thebibliography}{99}
\bibitem{Biro:1993qt}
T. S. Birò et al.,
\emph{Parton equilibration in relativistic heavy-ion collisions},
\href{https://doi.org/10.1103/PhysRevC.48.1275}
{\emph{Phys. Rev. C} \textbf{48} (1993) 1275}
[{\tt nucl-th/9303004}].

\bibitem{Florkowski:2012as}
W. Florkowski et al.,
\emph{Hydrodynamics of anisotropic quark and gluon fluids},
\href{https://doi.org/10.1103/PhysRevC.87.034914}
{\emph{Phys. Rev. C} \textbf{87} (2013) 034914}
[{\tt nucl-th/arXiv:1209.3671}].

\bibitem{Strickland:2018ayk}
M. Strickland,
\emph{The non-equilibrium attractor for kinetic theory in relaxation time approximation},
\href{https://doi.org/10.1007/JHEP12(2018)128}
{\emph{JHEP} \textbf{12} (2018) 128}
[{\tt nucl-th/arXiv:1809.01200}].

\bibitem{Florkowski:2013lya}
W. Florkowski et al.,
\emph{Testing viscous and anisotropic hydrodynamics in an exactly solvable case},
\href{https://doi.org/10.1103/PhysRevC.88.024903}
{\emph{Phys. Rev. C} \textbf{88} (2013) 024903}
[{\tt nucl-th/arXiv:1305.7234}].

\bibitem{Broniowski:2008qk}
W. Broniowski et al.,
\emph{Free-streaming approximation in early dynamics of relativistic heavy-ion collisions},
\href{https://doi.org/10.1103/PhysRevC.80.034902}
{\emph{Phys. Rev. C} \textbf{80} (2009) 034902}
[{\tt nucl-th/arXiv:0812.3393}].

\bibitem{Alqahtani:2017mhy}
M. Alqahtani et al.,
\emph{Relativistic anisotropic hydrodynamics},
\href{https://doi.org/10.1016/j.ppnp.2018.05.004}
{\emph{Prog. Part. Nucl. Phys} \textbf{101} (2018) 204}
[{\tt nucl-th/arXiv:1712.03282}].

\bibitem{Strickland:2017kux}
M. Strickland et al.,
\emph{Anisotropic non-equilibrium hydrodynamic attractor},
\href{https://doi.org/10.1103/PhysRevD.97.036020}
{\emph{Phys. Rev. D} \textbf{97} (2018) 036020}
[{\tt nucl-th/arXiv:1709.06644}].

\bibitem{Frasca:2024ege}
F. Frascà et al.,
\emph{Far-from-equilibrium attractors in kinetic theory for a mixture of quark and gluon fluids},
[{\tt hep-ph/arXiv:2407.17327}].

\end{thebibliography}
\end{document}